%% file: root.tex
\documentclass[letterpaper, 10 pt, conference]{ieeeconf}  % Comment this line out if you need a4paper

\IEEEoverridecommandlockouts                              % This command is only needed if 
\usepackage{graphicx}
\usepackage{dirtytalk}
\usepackage{etoolbox}
\usepackage{booktabs}
\usepackage{multirow}

\usepackage[mathscr]{euscript}

\usepackage[table,xcdraw]{xcolor}

\usepackage{hyperref}

\usepackage{amsmath}
\usepackage{amssymb}

\usepackage{algorithm}
\usepackage{algpseudocode}
\usepackage{enumerate}

\usepackage{threeparttable}
\usepackage{scalerel}
\usepackage{tikz}
\usepackage{cite}

\title{\LARGE \bf
Color-Quality Invariance for Robust Medical Image Segmentation
}

\author{Ravi Shah, Atsushi Fukuda, and Quan Huu Cap
\thanks{All authors are with the AI Development Department, Aillis, Inc., Tokyo, Japan. Corresponding author: Quan Huu Cap ({\tt\small quan.cap@aillis.jp})}
}

\begin{document}

\maketitle
\thispagestyle{empty}
\pagestyle{empty}

%%%%%%%%%%%%%%%%%%%%%%%%%%%%%%%%%%%%%%%%%%%%%%%%%%%%%%%%%%%%%%%%%%%%
\begin{abstract}
    \input{00_Abstract}
% \newline

% \indent \textit{Clinical relevance}— This is a brief additional statement on why a this might be of interest to practicing clinicians. Example: This establishes the anesthetic efficacy of 10\% intraosseous injections with epinephrine to positively influence cardiovascular function.
\end{abstract}

% KEYWORDS

%%%%%%%%%%%%%%%%%%%%%%%%%%%%%%%%%%%%%%%%%%%%%%%%%%%%%%%%%%%%%%%%%%%%%%%%%%%%%%%%
\section{INTRODUCTION}
    \input{01_Introduction}

\section{METHODS}
    \input{02_Methods}

\section{EXPERIMENTAL RESULTS}
    \input{03_Experimental_Results}

\section{DISCUSSION}
    \input{04_Discussion}

\section{CONCLUSION}
    \input{05_Conclusion}

\section*{ACKNOWLEDGMENT}
We would like to thank all researchers at Aillis Inc., especially the AI Development team, Dr. Memori Fukuda (MD), Dr. Kei Katsuno (MD), and Dr. Takaya Hanawa (MD) for their valuable comments and feedback. 

\nocite{*}
\footnotesize{
\bibliographystyle{IEEEtran}
\bibliography{reference}
}

\end{document}

%% file: 00_Abstract.tex
Single-source domain generalization (SDG) in medical image segmentation remains a significant challenge, particularly for images with varying color distributions and qualities. 
Previous approaches often struggle when models trained on high-quality images fail to generalize to low-quality test images due to these color and quality shifts. 
In this work, we propose two novel techniques to enhance generalization: dynamic color image normalization (DCIN) module and color-quality generalization (CQG) loss. 
The DCIN dynamically normalizes the color of test images using two reference image selection strategies. 
Specifically, the DCIN utilizes a global reference image selection (GRIS), which finds a universal reference image, and a local reference image selection (LRIS), which selects a semantically similar reference image per test sample. 
Additionally, CQG loss enforces invariance to color and quality variations by ensuring consistent segmentation predictions across transformed image pairs. 
Experimental results show that our proposals significantly improve segmentation performance over the baseline on two target domain datasets, despite being trained solely on a single source domain. 
Notably, our model achieved up to a 32.3-point increase in Dice score compared to the baseline, consistently producing robust and usable results even under substantial domain shifts. 
% Notably, our full DCIN pipeline (GRIS + LRIS) achieves superior generalization compared to existing methods, including those using human-expert-selected references. 
Our work contributes to the development of more robust medical image segmentation models that generalize across unseen domains.
The implementation code is available at \url{https://github.com/RaviShah1/DCIN-CQG}.

%% file: 01_Introduction.tex
Accurate segmentation of medical images is crucial for computer-aided diagnosis, healthcare systems, and clinical research. 
In recent years, with the rapid development of deep learning techniques, the performance of medical image segmentation tasks has been significantly improved \cite{qureshi2023medical, azad2024medical}. 
However, these deep learning-based models tend to perform well when trained and tested on datasets from a single domain but perform poorly when faced with data from different domains, which is known as the domain shift problem. 
To improve the generalization of these models on new test cases, we could use larger datasets that include a diverse range of high-quality images. 
However, obtaining such extensive medical datasets is very challenging, primarily due to the obstacles in data collection and sharing, as well as the high expenses associated with labeling. 

One solution to the problem of domain shift is to use unsupervised domain adaptation (UDA) techniques. 
In the UDA setting, segmentation models are trained on labeled data from a source domain, as well as unlabeled data from the target domain, in order to better generalize to the target domain \cite{guan2021domain, kumari2024deep}. 
However, a drawback of this method is that the target domain must be known in advance, which may not be possible in medical applications. 
Additionally, if the target domain changes, a new model would need to be re-trained, which can be difficult or impossible to accomplish in real-world scenarios. 

Another potential approach for overcoming the above issue is to employ domain generalization (DA) methods \cite{matta2024systematic, yoon2024domain}. 
These methods are designed to improve generalization by training only from labeled source domains without accessing data from unseen target domains. 
The DA methods could be divided into two categories: multi-source domain generalization (MDG) and single-source domain generalization (SDG). 
The former, MDG methods, normally require training on two or more labeled source domains. 
However, as previously noted, this is problematic due to the limited availability of medical images from various domains in real-world settings. 
On the other hand, the SDG methods are known to be more practical as these methods use training data from only one source domain but generalize it to unseen target data from multiple domains. 
Through these SDG methods, significant improvements in performance have been observed in the medical image segmentation tasks \cite{su2023rethinking,hu2023devil,yang2024single,liu2024universal,jiang2025structure}. 
% Fig. 1
\input{figures/tex_files/Fig_1}

Despite achieving promising results, they largely rely on the assumption of minimal domain distribution shifts. 
In practice, large domain shifts, particularly with color medical images, can lead to a noticeable decline in the performance of SDG segmentation models. 
Fig. \ref{fig:fig_1} depicts instances of failure from a throat segmentation model which was trained on high-quality (HQ) color throat images captured from a professional camera but tested on low-quality (LQ) images acquired from other camera models. 
While performing well on the HQ domain (first row), the model was unable to capture semantic information from other domains (second to last rows). 
In practice, failure cases are more likely to occur since LQ color medical images (e.g., low-resolution, noise, etc.,) such as endoscopic throat images are often obtained with very complex degradations during the acquisition process \cite{cap2025practical}. 
As we experienced, this decline in segmentation performance is mainly due to the differences in color between domains. In addition, previous studies have shown that the decisions of deep networks are greatly influenced by the color differences between datasets \cite{afifi2019else,de2021impact}. 
Therefore, we believe the implementation of advanced techniques for color distribution alignment has the potential to improve the segmentation performance. 

Color normalization is a common technique to mitigate the effects of domain shift in medical imaging \cite{macenko2009method, vahadane2016structure}. 
However, from our observations, segmentation results vary depending on the chosen reference images, making the selection of suitable reference images for testing a challenging task. 
While expert-selected reference images may provide guidance, this process remains highly subjective and dependent on individual judgment. 
In this work, we first propose a dynamic color image normalization (DCIN) method that dynamically selects suitable reference images from the training data and transfers the color distribution of test images to match that of the reference image. 
We show that the performance of segmentation models is better when using our DCIN module compared to expert reference image selection. 

To further enhance the robustness of the segmentation models, we introduce a training objective function called color-quality generalization (CQG) as our second proposal. 
The CQG loss is a contrastive-based loss with the idea that an image in different color and quality conditions should have the same segmentation outputs. 
These two proposals effectively mitigate the issue of large domain shifts and significantly improve the performance of SDG segmentation models. 
Experiments demonstrate that our proposed method results in large increases in DICE scores across all segmentation models on two other domain datasets, with a maximum improvement of 32.3 points over the baseline. 
Moreover, our method is model agnostic which can be adapted to any SGD color image segmentation model.  
% Fig. 2
\input{figures/tex_files/Fig_2}

%% file: figures/tex_files/Fig_1.tex
%Figure 1
\begin{figure}[!t]
\centering
\includegraphics[width=0.99\linewidth]{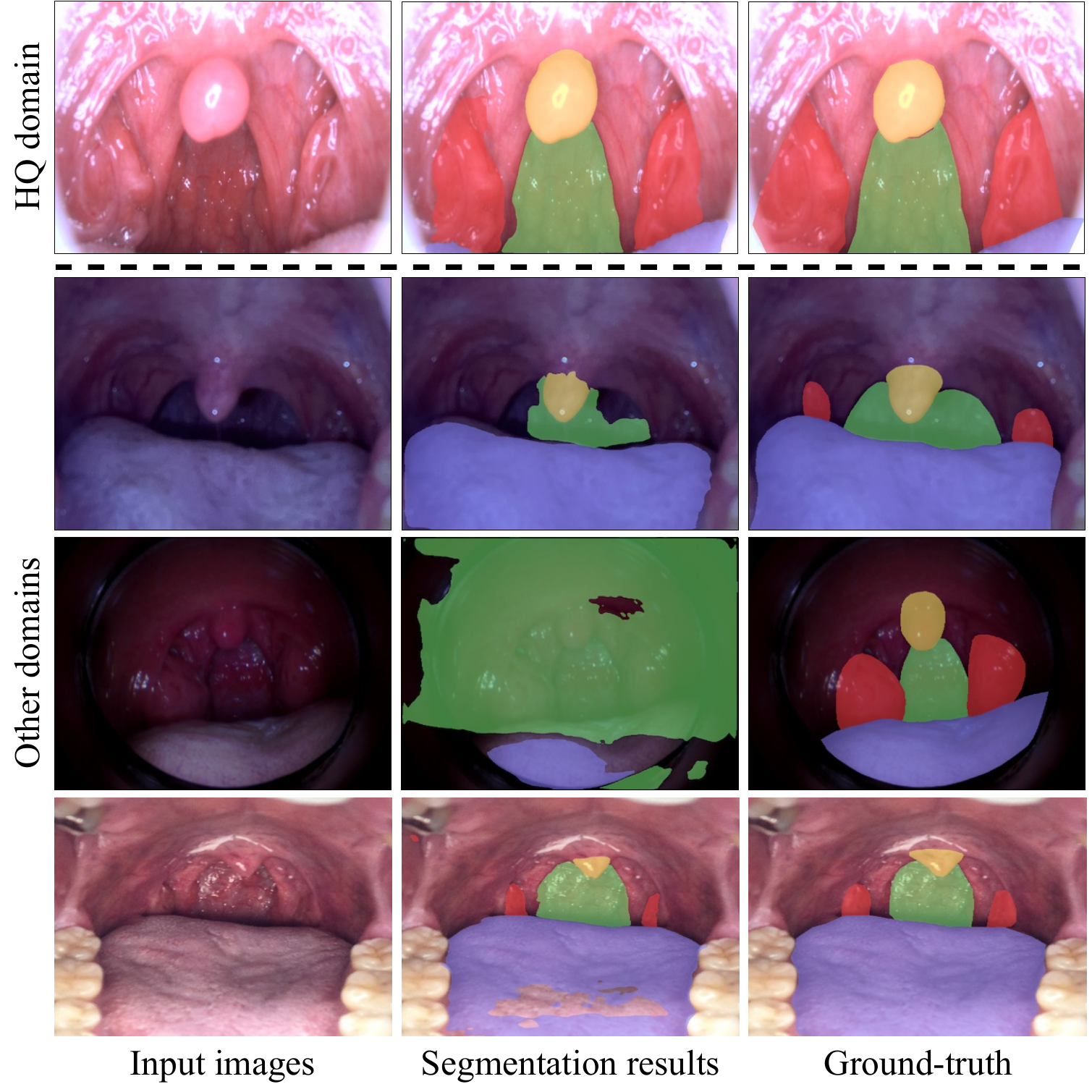}
\caption{
    Comparison of segmentation results on HQ domain (source domain) and LQ domain (other domains). 
    While performing well on the HQ domain (first row), the model was unable to capture semantic information from other domains (second to last rows). 
}
\label{fig:fig_1}
\end{figure}

%% file: figures/tex_files/Fig_2.tex
%Figure 2
\begin{figure*}[!t]
\centering
\includegraphics[width=0.8\textwidth]{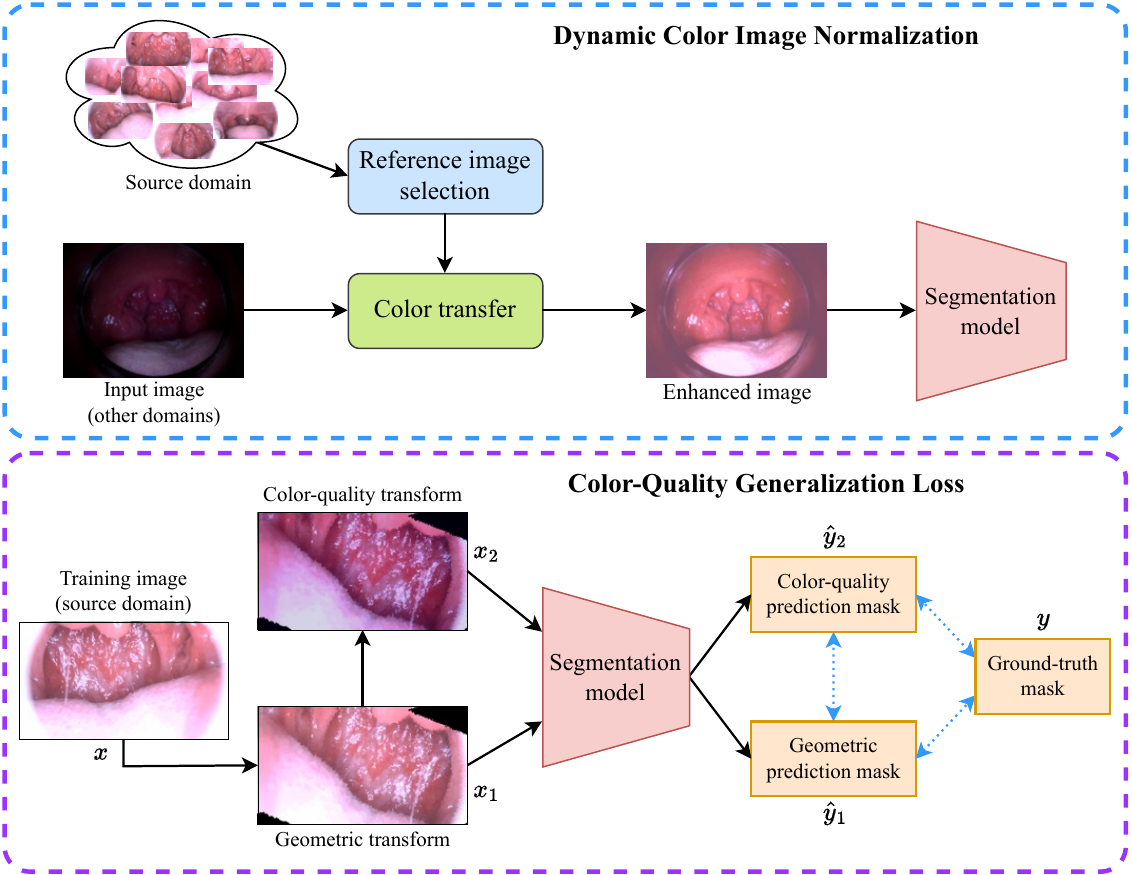}
\caption{
    The overview of our proposed dynamic color image normalization (DCIN) and color-quality generalization (CQG) loss.
}
\label{fig:fig_2}
\end{figure*}

%% file: 02_Methods.tex
\subsection{Dynamic Color Image Normalization}
Fig. \ref{fig:fig_2} (blue dashed box) shows the data flow of our dynamic color image normalization (DCIN) method. 
For a given input test image from a non-source domain, the reference image selection module strategically identifies suitable reference images from the training source domain. 
The color transfer in the perception-based color space $l\alpha \beta$ \cite{reinhard2001color} is then applied to align the color distribution of the reference images with that of the test image. 
The reference image selection module incorporates two strategies: \say{global} reference selection and \say{local} reference selection. 
The global strategy assigns a single reference image to all test images, whereas the local strategy selects a unique reference image for each individual test image. 
Detailed descriptions of these strategies are described below. 

\subsubsection{Global Reference Image Selection}
We propose the global reference image selection (GRIS) strategy that utilizes color histograms such that each image in the source domain is converted into a $b$-bin normalized color histogram vector. 
The global reference image $x_g$ is selected as the image whose color histogram minimizes the average pairwise distance between histograms of all other images in the source domain. 
The average pairwise distance $\mathcal{D}_{\mathrm{pairwise}}(x_i) = \frac1N\overset N{\underset{j=1}{\sum d_{i,j}}}$, where $d_{i,j}=\sqrt{\sum_{k=1}^b\left(H_k\left(x_i\right)-H_k\left(x_j\right)\right)^2}$ is the Euclidean distance between two normalized histogram vectors. 
In this case, $H_k(x)$ is the $k$-th bin value of the image $x$, and $N$ is the number of images in the source domain. 
The selected image $x_g$ will be used as the color normalization reference of \textit{all} test images before making predictions. 

\subsubsection{Local Reference Image Selection}
As stated previously, segmentation results vary on different reference images. 
Thus, we believe selecting a semantically similar image to the test image from the training data can benefit the segmentation performance. 
We propose a local reference image selection (LRIS) strategy that utilizes a pre-trained CNN model to select a more tailored reference image for each test image. 
First, the pre-trained CNN was used to extract feature vectors from all source images, which were then normalized into unit vectors. 
For a given image $x_{test}$ in a test domain, a local reference image $x_l$ is selected as the one whose feature vector has the highest cosine similarity with $x_{test}$. 
The selected image $x_l$ will be used as the color normalization reference of the test image $x_{test}$ before making predictions. 

\subsubsection{Ensembling Both Selected Reference Images}
To further improve generalization capabilities, the results of the above two strategies can be ensembles. 
From the selected global and local reference images, two color-normalized input images were produced, which were then used to generate two corresponding output masks. 
Finally, the final prediction is formed by taking a pixel-wise mean of the two predicted masks. 
Here, we refer to the \say{full DCIN} as the complete module with this ensemble method (i.e., GRIS + LRIS). 

\subsection{The Color-Quality Generalization Loss}
The color-quality generalization loss (CQG) is a training objective function inspired by contrastive loss. 
The idea is that given the same input presented in varying colors and qualities, the model should produce identical segmentation masks. 
This approach encourages the model to adapt effectively to images with diverse variations. 
Fig. \ref{fig:fig_2} (purple dashed box) illustrates the data flow of the CQG loss. 
For each training image $x$, we apply transformations randomly to generate two inputs for the segmentation model. 
The first input $x_1$ is obtained by applying geometric transformations, and the second input $x_2$ is from both geometric and photometric transformations. 
Geometric transformations are applied to change the input $x$ geometrically while photometric transformations change its color and quality. 
Specifically, geometric transformations include random horizontal flip, shear, shift, scale, rotation, and elastic transform. 
Photometric transformations include random blur, sharpening, Gaussian noise, brightness contrast, and RGB shifts. 
Both $x_1$ and $x_2$ have the same ground-truth mask $y$.

Given a segmentation model $S$ and a ground-truth mask $y$, we have $y_1 = S(x_1)$, and $y_2 = S(x_2)$. 
Our CQG loss is defined as:
\begin{equation}
    \mathcal{L} = \lambda_{1}\mathrm{DC}(y,y_1) + \lambda_{2}\mathrm{DC}(y,y_2) + \lambda_{3}\mathrm{MSE}(y_1,y_2),
\end{equation}
where $\mathrm{DC}(y,y\prime)$ is the sum of the Dice loss and cross-entropy loss between the ground truth $y$ and predicted mask $y\prime$. 
$\mathrm{MSE}(y_1,y_2)$ is the mean squared error loss between the predicted masks $y_1$ and $y_2$. 
Here, $\lambda_{1}, \lambda_{2}, \lambda_{3}$ are the hyper-parameters controlling the weight of each loss term. 

The CQG loss can be viewed as a form of image augmentation. By leveraging this loss, the model is encouraged to produce identical predictions for both original and augmented images, regardless of color and quality shifts. 

%% file: 03_Experimental_Results.tex
\subsection{Data Collection}
In this work, we collected 16,000 high-quality (HQ) clean throat images of around 900 patients from several hospitals in Japan and refer to it as the $\mathrm{HQ}_{all}$ dataset. 
They were obtained by a special camera type designed for taking pharyngeal images. 
Among the $\mathrm{HQ}_{all}$ dataset, there are 2,000 images that were labeled by experts containing the semantic pixel-level annotations of four areas inside the throat: uvula, tongue, tonsil, and pharyngeal wall. 
We randomly split 1,600 images for training (refer as $\mathrm{HQ}_{seg/train}$ set) and 400 images for validation (refer as $\mathrm{HQ}_{seg/val}$ set). 
In addition, the $\mathrm{HQ}_{all}$ dataset is used for color reference image selection as in the DCIN module. 

To evaluate the performance of the medical image segmentation models, we additionally collected data from two non-source domains. 
First, we collected 255 LQ images (e.g., blurry, hazy, compression artifacts, poor lighting, etc.) from over 150 patients. 
These images were taken by different camera devices, and we refer this as the $\mathrm{LQ}_{seg}$ dataset. 
Second, we collected 125 smartphone (SP) images. 
These images were taken primarily with iPhone SE (1st gen) and Sony Xperia XZ1 (SO-01K) cameras, and we refer to this as the $\mathrm{SP}_{seg}$ dataset. 
Images from all datasets are resized to the size of $768 \times 512$ pixels. 
Please refer to Fig. \ref{fig:fig_1} for some samples of these datasets. 

\subsection{Reference Image Selection}
For each test image, we utilize two images that correspond to the GRIS and LRIS strategies in our DCIN module. 
The global reference image $x_g$ is selected by applying the color histogram reference image selection pipeline on the $\mathrm{HQ}_{seg/train}$ dataset. 
We created the histograms in RGB space and used eight bins per channel based on our preliminary experiments. 
Note again that the image is the same across all test images and is selected before the test time. 
The local reference image $x_l$ is selected at inference time for each test image. 
Specifically, for all images in the $\mathrm{HQ}_{all}$ dataset, we pre-computed embedding vectors using a Swin-V2-Large model \cite{liu2021swin} which was pre-trained on the ImageNet dataset \cite{deng09imagenet}. 

To further validate the effectiveness of our GRIS and LRIS strategies, we incorporated a reference image selected by physicians for color transfer before testing. 
The image was chosen based on specific criteria, including cleanliness and color balance, ensuring its suitability for diagnostic purposes. 
We refer to this image as the expert-selected reference image (ExRI). 
% Fig. 3
\input{figures/tex_files/Fig_3}

\subsection{Training Throat Segmentation Models}
To evaluate the effectiveness of our proposals for supporting SDG medical image segmentation tasks, in this experiment, we trained three different throat image segmentation models on the $\mathrm{HQ}_{seg/train}$ dataset. 
All models are built based on the U-Net \cite{ronneberger15unet} model with the pre-trained EfficientNet-B2 \cite{tan2019efficientnet} on the ImageNet dataset as the backbone. 
The three segmentation models are:
\begin{itemize}
    \item Baseline ($S_{\mathrm{base}}$): The baseline model is trained with minimal preprocessing, consisting only of resizing and normalization. 
    A simple loss function comprising the sum of Dice and cross-entropy losses is applied without any data augmentations. 
    
    \item Baseline + augmentations ($S_{\mathrm{aug}}$): The baseline model is additionally trained with the augmentations as in the CQG loss. 
    The data augmentation only creates one output from an input image and the CQG loss is not applied. 
    
    \item Baseline + CQG ($S_{\mathrm{CQG}}$): The baseline model is trained with the color-quality generalization (CQG) loss function. 
    We chose to use $\lambda_{1}=0.3, \lambda_{2}=0.7, \lambda_{3}=1.0$ based on our preliminary experiments. 
\end{itemize}

All three throat segmentation models were optimized using the Adam optimizer \cite{kingma2015adam} with a learning rate of $1 \times 10^{-3}$. 
The batch size was set to 2 and training was completed after 15 epochs. 
For the evaluation metric, we employed the commonly used Dice score to measure the overlap between the prediction and the segmentation ground-truth masks. 

\subsection{Results of Throat Segmentation Models}
After training, the three segmentation models $S_{\mathrm{base}}$, $S_{\mathrm{aug}}$, and $S_{\mathrm{CQG}}$ achieved Dice scores of 88.9, 87.8, and 88.6 on the $\mathrm{HQ}_{seg/val}$ dataset, respectively. 
All models demonstrated high accuracy on the source domain, indicating the capability of accurately segmenting HQ images. 

Table \ref{tab:table_1} and \ref{tab:table_2} summarize the Dice scores for the three models on the $\mathrm{LQ}_{seg}$ and $\mathrm{SP}_{seg}$ datasets (the best performance of each model across all DCIN configurations is in \textbf{bold} text). 
The visual comparison of segmentation results is provided in Fig. \ref{fig:fig_3}. 
Note that the results with DCIN depicted in Fig. \ref{fig:fig_3} are from the full DCIN (i.e., GRIS + LIRS). 
All models experienced a severe performance drop when moving from the HQ dataset to other non-source datasets. 
For instance, without DCIN module, the $S_{\mathrm{base}}$ model's Dice score largely dropped from 88.9 on $\mathrm{HQ}_{seg/val}$ to 36.9 on the $\mathrm{LQ}_{seg}$ dataset, and down to 55.3 on the $\mathrm{SP}_{seg}$ dataset. 

Both the DCIN module and CQG loss consistently boosted the segmentation performances in all models. 
The most effective configuration was achieved by combining CQG training with the full DCIN (i.e., GRIS + LRIS), resulting in increments in Dice score over the baseline models, from 36.9 to 69.2 on the $\mathrm{LQ}_{seg}$, and 55.3 to 68.6 on the $\mathrm{SP}_{seg}$ datasets. 
% Table I
\input{tables/Table_I}
% Table II
\input{tables/Table_II}

%% file: figures/tex_files/Fig_3.tex
%Figure 3
\begin{figure*}[!t]
\centering
\includegraphics[width=0.95\textwidth]{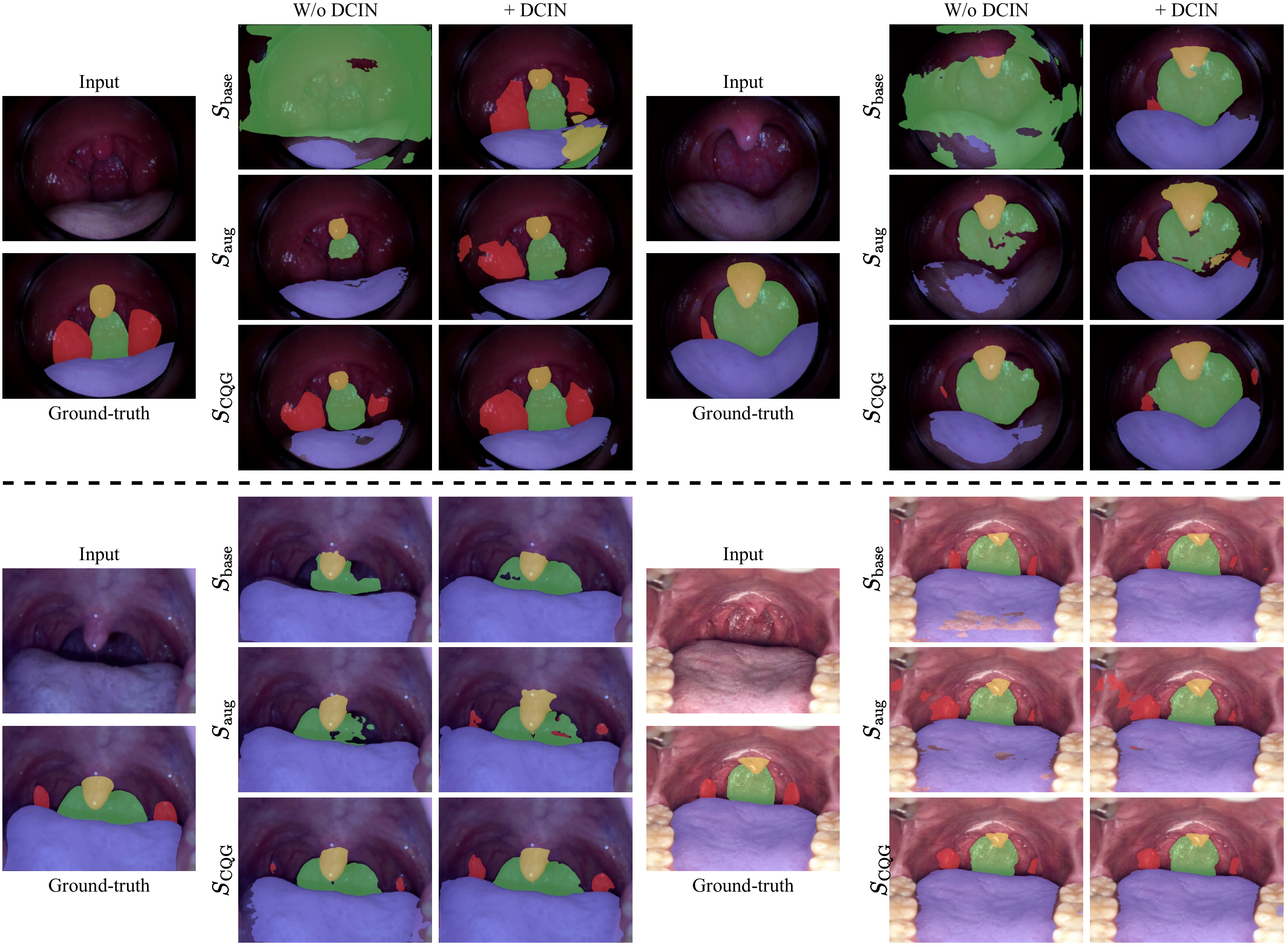}
\caption{
    Visual comparison of the results on the $\mathrm{LQ}_{seg}$ (top) and $\mathrm{SP}_{seg}$ (bottom) datasets from different segmentation models. 
    Results with DCIN are from the full DCIN setting (i.e., GRIS + LRIS).
}
\label{fig:fig_3}
\end{figure*}

%% file: tables/Table_I.tex
\begin{table}[t]
\centering
\caption{Performance comparison in Dice score on the $\mathrm{LQ}_{seg}$ dataset}
\label{tab:table_1}
\resizebox{0.97\linewidth}{!}{
\begin{tabular}{@{}llllll@{}}
\toprule
\multicolumn{1}{c}{\multirow{2}{*}{Model}} & \multirow{2}{*}{W/o DCIN} & \multicolumn{4}{c}{DCIN}                    \\ \cmidrule(l){3-6} 
\multicolumn{1}{c}{}                        &                           & ExRI & GRIS & LRIS          & GRIS + LRIS   \\ \midrule
$S_{\mathrm{base}}$                                    & 36.9                      & 53.7 & 52.4 & \textbf{58.9} & 56.9          \\ \midrule
$S_{\mathrm{aug}}$                                & 54.0                      & 55.9 & 53.7 & \textbf{58.1} & 57.2          \\ \midrule
$S_{\mathrm{CQG}}$                                & 64.5                      & 67.2 & 67.5 & 68.5          & \textbf{69.2} \\ \bottomrule
\end{tabular}
}
\end{table}

%% file: tables/Table_II.tex
\begin{table}[t]
\centering
\caption{Performance comparison in Dice score on the $\mathrm{SP}_{seg}$ dataset}
\label{tab:table_2}
\resizebox{0.97\linewidth}{!}{
\begin{tabular}{@{}llllll@{}}
\toprule
\multicolumn{1}{c}{\multirow{2}{*}{Model}} & \multirow{2}{*}{W/o DCIN} & \multicolumn{4}{c}{DCIN}                    \\ \cmidrule(l){3-6} 
\multicolumn{1}{c}{}                       &                           & ExRI & GRIS & LRIS          & GRIS + LRIS   \\ \midrule
$S_{\mathrm{base}}$                                   & 55.3                      & 61.0 & 63.8 & 60.8 & \textbf{65.5} \\ \midrule
$S_{\mathrm{aug}}$                               & 60.6                      & 62.4 & 66.1 & 64.4 & \textbf{66.7} \\ \midrule
$S_{\mathrm{CQG}}$                               & 63.5                      & 65.0 & 68.1 & 67.1          & \textbf{68.6} \\ \bottomrule
\end{tabular}
}
\end{table}

%% file: 04_Discussion.tex
% In this paper, we studied and proposed a training technique for model robustness and a dynamic image enhancement pipeline for supporting the SDG medical color image segmentation tasks. 
Due to the large domain shifts between different throat datasets, the $S_{\mathrm{base}}$ model and the $S_{\mathrm{aug}}$ model (even with data augmentations), struggled to generalize on the unseen images, resulting in substantial performance drops on both the $\mathrm{LQ}_{seg}$ and $\mathrm{SP}_{seg}$ datasets. 
Visually, the results in Fig. \ref{fig:fig_3} illustrate that the baseline $S_{\mathrm{base}}$ model's predictions without DCIN are completely unusable. 
As discussed earlier, the large difference in image quality and image color between domains is the main reason for the vast decline in the performance of deep learning models. 

% We address these issues at both train time (using the CQG loss) and test time (using the DCIN module) and achieve greatly improved Dice scores for all segmentation models. 
We address these issues during both training (using the CQG loss) and testing (using the DCIN module). 
The CQG loss demonstrated its effectiveness as the $S_{\mathrm{CQG}}$ model outperformed other models in all experiments. 
By employing the contrastive constraint on different color and quality conditions from the same input, the CQG loss has increased the robustness of the model on other data domains. 
In addition, applying the DCIN module also significantly improved the segmentation performance on both datasets (Table \ref{tab:table_1} and \ref{tab:table_2}). 
The $S_{\mathrm{CQG}}$ with the full DCIN (GRIS + LRIS) produces results that closely resemble the ground truth. 
Notably, its visual results on the $\mathrm{SP}_{seg}$ dataset (Fig. \ref{fig:fig_3}, bottom part) appear very clean and accurate. 
In this context, the DCIN module effectively aligns the color distribution with that of the source domain, further boosting the segmentation performances. 

To confirm the impact of the two image reference selection strategies (GRIS and LRIS) in the DCIN module versus expert-selected reference image (ExRI), we also reported the results under different configurations: DCIN using ExRI, GRIS, and LRIS. 
In most experiments, the LRIS and GRIS strategies outperformed the DCIN with ExRI. 
This suggests that the subjective expert-selected reference image is not optimal while our objective GRIS and LRIS strategies often provided more suitable reference images. 

Although achieving promising results, there remain several limitations in our proposals. 
First, our current DCIN performed $l\alpha \beta$ color transfers on the CPU, which is somewhat inefficient as the image is later processed on the GPU. 
Adapting this operation to run directly on the GPU could improve processing speed. 
Second, the segmentation model faces challenges when target domains contain artifacts absent in the source domain. 
For instance, many images in the $\mathrm{SP}_{seg}$ dataset displayed teeth, which often confused the model and degraded prediction quality. 
Introducing additional post-processing techniques could improve the model’s ability to generalize when domain shifts extend beyond color and quality differences. 
There is room for improvement and we plan to mitigate these limitations in future works. 

%% file: 05_Conclusion.tex
In this study, we proposed a practical approach for single-source domain generalization in color medical image segmentation. 
By introducing dynamic color image normalization (DCIN) for test-time color adaptation and color-quality generalization (CQG) loss to enhance model robustness, our method effectively improves segmentation performance on non-source domains. 
Trained solely on high-quality pharynx images, our framework demonstrates strong generalization to low-quality and smartphone-acquired images. 
The proposed pipeline not only enhances segmentation accuracy across large domain shifts for pharyngeal images but also highlights its potential for broader applications to improve the robustness of medical imaging systems.